\documentclass[authoryear,preprint,11pt,a4paper]{elsarticle}

\usepackage{color,soul}
\usepackage{graphics}
\usepackage{epsfig}
\usepackage{rotating}
\usepackage{amssymb}

\usepackage{supertabular}

\usepackage{longtable}

\usepackage{lscape}

\usepackage{fullpage}

\usepackage{lineno}

\journal{Icarus}

\begin{document}

\begin{frontmatter}





\title{Olivine-dominated asteroids: Mineralogy and origin}

\author[MPS]{Juan A. Sanchez\corref{cor1}\fnref{fn1}}
\ead{sanchez@mps.mpg.de}

\author[PSI]{Vishnu Reddy\fnref{fn1}}

\author[GSU]{Michael S. Kelley\fnref{fn1,fn2}}

\author[HOSERLAB]{Edward A. Cloutis}

\author[SRI]{William F. Bottke}

\author[SRI]{David Nesvorn\'{y}}

\author[UT]{Michael P. Lucas}

\author[UND]{Paul S. Hardersen\fnref{fn1}}

\author[UND]{Michael J. Gaffey\fnref{fn1}}

\author[NASA]{Paul A. Abell\fnref{fn1}}

\author[PSI]{Lucille Le Corre}

\address[MPS]{Max Planck Institut f$\ddot{u}$r Sonnensystemforschung, Katlenburg-Lindau, Germany}

\address[PSI]{Planetary Science Institute, 1700 East Fort Lowell Road, Tucson, Arizona 85719, USA}

\address[GSU]{Department of Geology and Geography, Georgia Southern University, Statesboro, USA}

\address[HOSERLAB]{Department of Geography, University of Winnipeg, Winnipeg, Manitoba, Canada}

\address[SRI]{Southwest Research Institute and NASA Lunar Science Institute, Boulder, USA}

\address[UT]{Department of Earth and Planetary Sciences, University of Tennessee, USA}

\address[UND]{Department of Space Studies, University of North Dakota, Grand Forks, USA}

\address[NASA]{NASA Johnson Space Center, Houston, Texas, USA}

\fntext[fn1]{Visiting Astronomer at the Infrared Telescope Facility, which is operated by the University of Hawaii under Cooperative Agreement No. NNX-08AE38A with the National Aeronautics and Space Administration, Science Mission Directorate, Planetary Astronomy Program.}

\fntext[fn2]{Planetary Science Division, Science Mission Directorate, NASA Headquarters, Washington, DC 20546, USA.}

\cortext[cor1]{Corresponding author at: Max Planck Institut f$\ddot{u}$r Sonnensystemforschung, Max Planck Str.2, 37191 Katlenburg-Lindau, Germany.}


\begin{abstract}

Olivine-dominated asteroids are a rare type of objects formed either in nebular processes or through magmatic differentiation. The analysis of meteorite samples suggest that at least 100 parent bodies in the main belt 
experienced partial or complete melting and differentiation before being disrupted. However, only a few olivine-dominated asteroids, representative of the mantle of disrupted differentiated bodies, are known to 
exist. Due to the paucity of these objects in the main belt their origin and evolution have been a matter of great debate over the years. In this work we present a detailed mineralogical analysis of twelve olivine-dominated 
asteroids. We have obtained near-infrared (NIR) spectra (0.7 to 2.4 $\mu$m) of asteroids (246) Asporina, (289) Nenetta, (446) Aeternitas, (863) Benkoela, (4125) Lew Allen and (4490) Bamberry. Observations were 
conducted with the Infrared Telescope Facility (IRTF) on Mauna Kea, Hawai'i. This sample was complemented with spectra of six other olivine-dominated asteroids including (354) Eleonora, (984) Gretia, (1951) Lick, 
(2501) Lohja, (3819) Robinson and (5261) Eureka obtained by previous workers. Within our sample we distinguish two classes, one that we call monomineralic-olivine asteroids, which are those whose spectra only exhibit 
the 1 $\mu$m feature, and another referred to as olivine-rich asteroids, whose spectra exhibit the 1 $\mu$m feature and a weak (Band II depth $\sim$ 4\%) 2 $\mu$m feature. For the monomineralic-olivine asteroids the 
olivine chemistry was found to range from $\sim$ Fo$_{49}$ to Fo$_{70}$, consistent with the values measured for brachinites and R chondrites. In the case of the olivine-rich asteroids we determined their olivine and 
low-Ca pyroxene abundance using a new set of spectral calibrations derived from the analysis of R chondrites spectra. We found that the olivine abundance for these asteroids varies from 0.68 to 0.93, while the fraction of 
low-Ca pyroxene to total pyroxene ranges from 0.6 to 0.9. A search for dynamical connections between the olivine-dominated asteroids and asteroid families found no genetic link (of the type core-mantel-crust) between 
these objects.

\end{abstract}


\end{frontmatter}

\section{Introduction}

A-type asteroids are a unique class of objects that were initially distinguished from the R-type asteroids (the group into which they'd previously been classified) based on broadband spectrophotometry 
by \citet{1983Icar...55..177V} and were later re-classified based on Eight Color Asteroid Survey (ECAS) data (0.3 to 1.1 $\mu$m) by \citet{1984PhDT.........3T}. Asteroids of this taxonomic class have moderately high 
albedos, extremely reddish slopes shortward of 0.7 $\mu$m, and a strong absorption feature centered at $\sim$1.05 $\mu$m \citep{1989aste.conf..298T}. Subsequent near-infrared (NIR) spectra have shown that in 
these ÔoriginalÕ A-type asteroids a $\sim$2 $\mu$m feature is absent or very weak, consistent with a silicate component of nearly monomineralic olivine on the surface of these bodies.

The discovery of olivine-dominated asteroids is of considerable interest regarding the accretion and geochemical evolution of primitive bodies. Olivine-dominated objects are expected to form either through magmatic 
differentiation, being the major constituent of the mantles of most differentiated bodies \citep{1996M&PS...31..607B}, or through nebular processes which can produce olivine-dominated objects like the R-chondrite parent 
body \citep{1994Metic..29..275S}. The presence of olivine-dominated asteroids suggests that at least some objects in the asteroid belt underwent complete or near-complete melting that led to the differentiation of their 
interiors. Another interesting aspect is that in order for the mantle to be exposed, the parent body must be fragmented or its deep interior exposed by large impacts. Based on meteorites in terrestrial collections, it is 
estimated that at least $\sim$ 100 meteorite parent bodies should have existed in the asteroid belt that underwent partial or complete melting and differentiation before disruption and fragmentation 
\citep{2000P&SS...48..887K}. However, even assuming all A-type asteroids are olivine-dominated, only a handful of objects from the mantles of differentiated and disrupted parent bodies were discovered during the 
taxonomic surveys. This is described as the "missing mantle" problem because the corresponding mantle components of the iron cores (as represented by iron meteorites) are missing 
\citep{1986MmSAI..57..103C, 1989aste.conf..921B, 1996M&PS...31..607B}. More recent work on M-type asteroids by \citet{2011M&PS...46.1910H} indicates that a subset of that population (766 Moguntia, 798 Ruth and 
1210 Morosovia) shows a significant olivine component in the surface assemblage. It is unclear, at this point, if the olivine seen on these M-type asteroids are pieces of mantle remaining on an iron-rich core or formed via 
nebular processes.

More recent surveys like the Small Main-Belt Asteroid Spectroscopic Survey (SMASS) \citep{1995Icar..115....1X}, and SMASS II \citep{2002Icar..158..146B, 2002Icar..158..106B} have expanded the number of members 
within each taxonomic class based on visible spectroscopy. Twelve new A-type asteroids were added to the original five from \citet{1984PhDT.........3T}. \citet{2002Icar..159..468B} observed 10 
A-type asteroids at near-infrared wavelengths, four from \citet{1984PhDT.........3T} and six from \citet{2002Icar..158..146B, 2002Icar..158..106B}. They subsequently divided the A-type asteroids into two groups based on 
the strength of the 1 $\mu$m feature. Because the \citet{2002Icar..159..468B} data do not extend beyond 1.65 $\mu$m, the possibility of a 2 $\mu$m feature due to pyroxene cannot be ruled out. It is important to note that 
A-type asteroids under the \citet{2002Icar..158..146B, 2002Icar..158..106B} taxonomic system are not the same as those under the original Tholen system. Some A-types in 
the SMASS II taxonomic system contain up to $\sim$ 20\% pyroxene, as indicated by the presence of a 2 $\mu$m feature in NIR data, e.g., (4142) Dersu-Uzala \citep{2004P&SS...52..291B}, and are similar to S-I/S-II 
asteroids in the Gaffey S-asteroid subtypes \citep{1993Icar..106..573G}. 

A comprehensive summary of all previous work on A-type asteroids is published in Sunshine et al. (2007). This study was based on the work of  \citet{1998JGR...10313675S}, 
and included VIS-NIR spectra of nine olivine-dominated asteroids. Of these nine objects, four were analyzed using the Modified Gaussian Model (MGM) \citep{1990JGR....95.6955S} in order to derive their olivine 
compositions. Those four objects that included (1951) Lick, (289) Nenetta, (246) Asporina, and (354) Eleonora were characterized by the lack of a detectable 2 $\mu$m feature. The other five asteroids; (446) Aeternitas, 
(863) Benkoela, (984) Gretia, (2501) Lohja, and (3819) Robinson, whose spectra have a detectable 2 $\mu$m feature, were not analyzed due to the difficulties inherent to the modeling of olivine-pyroxene 
mixtures \citep{2007M&PS...42..155S}. 

In the present work we analyze VIS-NIR spectra of twelve olivine-dominated asteroids, six observed by our group and six obtained from previous studies. Because taxonomic classification can be ambiguous depending on 
the system used we will refer to these objects as S(I)-types, which is the designation introduced by \citet{1993Icar..106..573G} that includes objects where olivine is the major silicate phase present. We further distinguish 
two classes within our sample: one class that will be called {\it monomineralic-olivine} asteroids, which are those whose spectra exhibit the 1 $\mu$m feature and no detectable 2 $\mu$m feature, and another class that will 
be called {\it olivine-rich} asteroids, whose spectra exhibit the 1 $\mu$m feature and a weak 2 $\mu$m feature. 

The approach used in the present study differs from previous work \citep[e.g.,][]{1998JGR...10313675S, 2007M&PS...42..155S} in that olivine compositions are determined from the measured Band I centers, along with 
a spectral calibration derived from laboratory measurements. Furthermore, in the case of the olivine-rich asteroids, the olivine-pyroxene abundance ratio (ol/(ol+px)) and the ratio of low-Ca pyroxene (LCP) to total 
pyroxene (LCP/(LCP+HCP)) are determined using a set of equations derived from the analysis of meteorite samples. Here we define low-Ca pyroxenes (LCP) as pyroxenes with $<$25\% iron and include pigeonite 
and orthopyroxene, and high-Ca pyroxenes (HCP) as those with $>$25\% iron, including augite-diopside-hedenbergite.

In addition to the mineralogical analysis we also search for dynamical connections between the studied objects and asteroid families. Using this information we finally discuss possible formation scenarios for the 
olivine-dominated asteroids.

\section{Observations and data reduction}

Observations were carried out with the NASA IRTF on Mauna Kea, Hawai'i. NIR spectra ($\sim$ 0.7-2.5 $\mu$m) were obtained with the SpeX instrument \citep{2003PASP..115..362R} 
 in its low resolution (R$\sim$150) prism mode with a 0.8" slit width. A typical observing sequence consists of spectra taken in pairs (A-beam and B-beam) by nodding the telescope. Nodding enables the subtraction of 
 the sky background from the object during the data reduction process. Depending on the magnitude of the asteroid, 10-20 spectra are taken per asteroid with a maximum integration of 120 seconds due to 
 saturation from the background sky. In order to correct for telluric water vapor features and to obtain relative reflectance values, local standard and solar analog stars were also observed. For each night, flat fields and arc 
 line spectra were acquired. Data reduction was carried out with Spextool \citep{2004PASP..116..362C}. Detailed descriptions of observation and data reduction protocols are presented in \citet{Reddy09} and 
 \citet{2011Icar..216..184R, 2012Icar..219..382R}. NIR spectra are normalized to unity at 1.5 $\mu$m. Table \ref{t:Table1} lists observational circumstances for the observed asteroids.    

\begin{figure*}[!ht]
\begin{center}
\includegraphics[height=12cm]{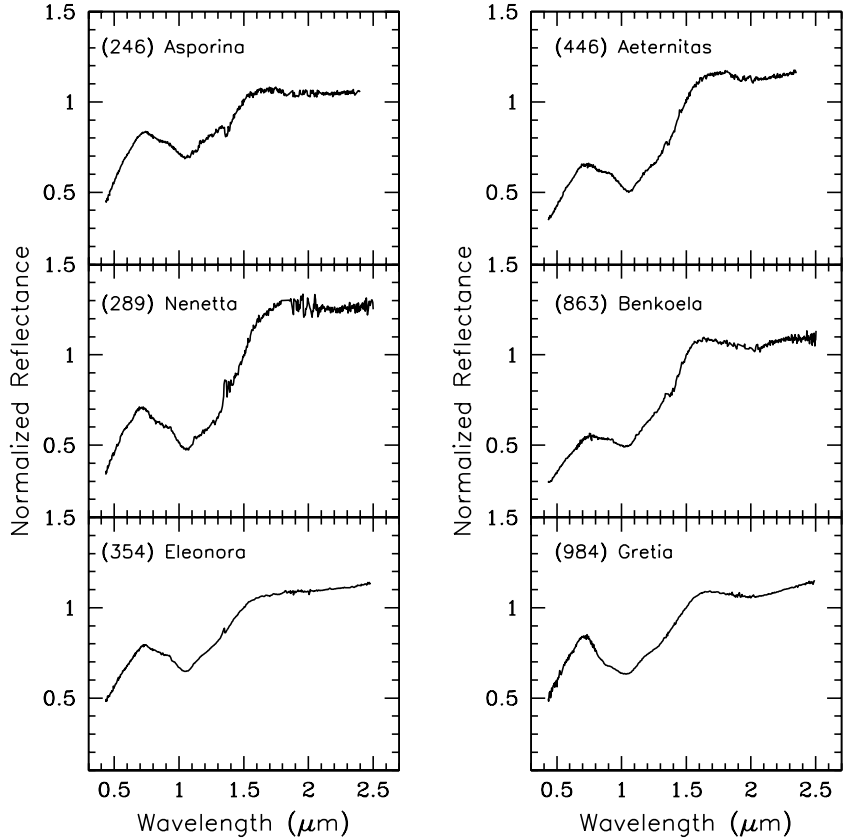}
\caption{\label{f:Aspec1} {\small VIS-NIR spectra of asteroids analyzed in this study. All spectra are normalized to unity at 1.5 $\mu$m. The noise seen in the spectrum of (289) Nenetta longwards of 1.9 $\mu$m is due 
to incompletely corrected telluric bands. The NIR spectrum of (354) Eleonora was obtained from the NEOSR survey (http://smass.mit.edu/minus.html). The NIR spectrum of (984) Gretia was obtained 
from \citet{2007M&PS...42..155S}. VIS spectra of all asteroids were obtained from the SMASS II survey \citep{2002Icar..158..146B, 2002Icar..158..106B}.}}
\end{center}
\end{figure*}

\begin{figure*}[!ht]
\begin{center}
\includegraphics[height=12cm]{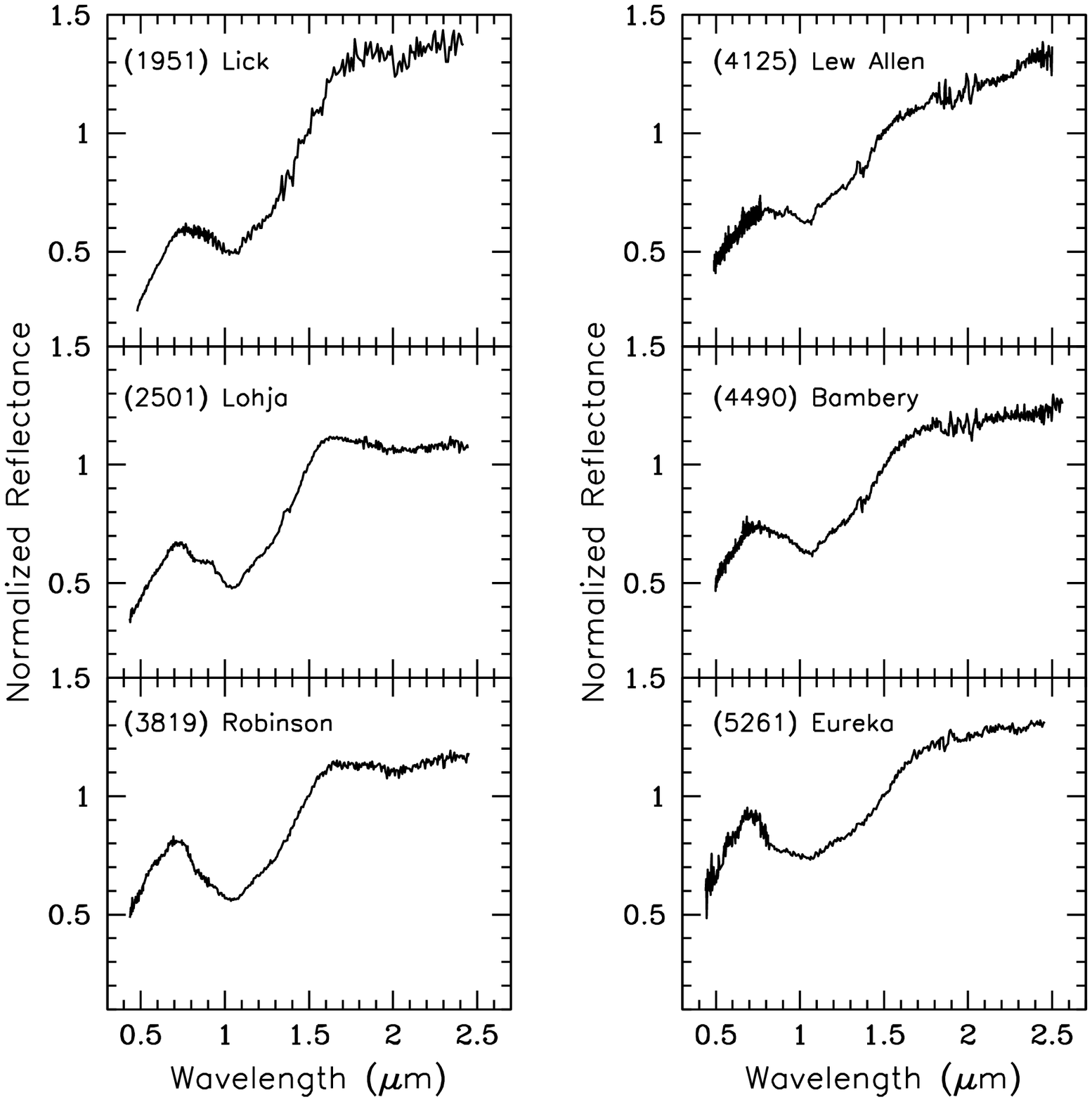}
\caption{\label{f:Aspec2} {\small VIS-NIR spectra of asteroids analyzed in this study. All spectra are normalized to unity at 1.5 $\mu$m. The noise seen in the spectra of (1951) Lick, (4125) Lew Allen and (4490) Bambery  
longwards of 1.9 $\mu$m is due to incompletely corrected telluric bands. The VIS-NIR spectrum of (1951) Lick was obtained from \citet{2010A&A...517A..23D}, the NIR spectrum of (2501) Lohja was obtained 
from \citet{2007M&PS...42..155S}, NIR spectra of (3819) Robinson and (5261) Eureka were obtained from the NEOSR survey (http://smass.mit.edu/minus.html). VIS spectra of asteroids (2501) Lohja, (3819) Robinson, 
and (5261) Eureka were obtained from the SMASS II survey  \citep{2002Icar..158..146B, 2002Icar..158..106B}. VIS spectra of asteroids (4125) Lew Allen and (4490) Bambery were obtained from 
the S$^{3}$OS$^{2}$ survey \citep{2004Icar..172..179L}.}}
\end{center}
\end{figure*}

In order to extend our study we also analyzed data from \citet{2007M&PS...42..155S}, \citet{2010A&A...517A..23D}, and the MIT-UH-IRTF Joint Campaign for NEO Spectral Reconnaissance (NEOSR). NIR spectra 
were combined with VIS spectra in order to increase the wavelength coverage. VIS spectra were obtained from the SMASS II \citep{2002Icar..158..146B, 2002Icar..158..106B}, and the Small Solar System Objects 
Spectroscopic Survey (S$^{3}$OS$^{2}$) \citep{2004Icar..172..179L}. VIS-NIR spectra of the studied asteroids are shown in Figs. \ref{f:Aspec1} and \ref{f:Aspec2}.

\begin{table}[!ht]
\caption{\label{t:Table1} {\small Observational circumstances. The columns in this table are: object number and designation, UT date, number of exposures, phase angle ($\alpha$), V-magnitude at the time of 
observation, heliocentric distance (r), air mass and solar analog used. The integration time with IRTF/SpeX was 120 s}.}
\begin{center}\small
\begin{tabular}{|c|c|c|c|c|c|c|c|}

\hline
Object&UT date&Exp&$\alpha$ ($^\mathrm{o}$)&mag. (V)&r (AU)&Air mass&Solar analog \\ \hline
246 Asporina&20-Jan-2004&10&5.5&15.0&2.88&1.16&Hyades 64 \\ 
289 Nenetta&17-Feb-2003&16&15.0&15.4&3.46&1.16&Hyades 64 \\
446 Aeternitas&20-Jan-2004&10&13.7&12.4&3.13&1.00&Hyades 64 \\ 
863 Benkoela&17-Nov-2002&6&17.9&14.8&3.20&1.02&Hyades 64 \\ 
4125 Lew Allen&15-Aug-2003&12&21.2&15.0&1.70&1.00&BS5996 \\
4490 Bamberry&17-Feb-2003&18&24.2&15.7&1.88&1.04&Hyades 64 \\

\hline

\end{tabular}
\end{center}
\end{table}


\clearpage

\section{Results}

\subsection{Monomineralic-olivine asteroids}

Asteroids in our sample that fall into the class of {\it monomineralic-olivines} are: (246) Asporina, (289) Nenetta, (354) Eleonora, (1951) Lick, (4125) Lew Allen, (4490) Bambery, and (5261) Eureka. The spectra 
of these asteroids lack detectable 2 $\mu$m feature (Band II depth $<$ 1\%), and the combined VIS-NIR spectra typically overlap at $\sim$ 0.7 $\mu$m. For each spectrum the Band I center and Band I depth are 
calculated after dividing out the linear continuum (a straight line tangent to the reflectance maxima) and fitting a polynomial over the bottom third of the band. A detailed explanation about the procedure used to measure 
the band parameters and their uncertainties can be found in \citet{2012Icar..220..36S, 2013Icar..225..131S}. The band parameters with their corresponding errors are presented in Table \ref{t:Table2}. 

\begin{table}[!ht]
\caption{\label{t:Table2} {\small Spectral band parameters for the monomineralic-olivine asteroids. The columns in this table correspond to: object number and designation, 
Band I center (BIC$\pm$0.005), temperature corrected Band I center ($\Delta$BIC$\pm$0.005), olivine composition (Fo$\pm$5\%) and its temperature corrected 
value ($\Delta$Fo$\pm$5\%), Band I depth (BI$_{\rm{depth}}\pm$0.3), geometric albedo ($p_{v}$), the beaming parameter ($\eta$) and the average surface temperature (T). Temperature values were calculated in the 
same way as \citet{2009M&PS...44.1331B}. The infrared emissivity, $\varepsilon$ is assumed to be 0.9, albedo and $\eta$ values of asteroids (246) Asporina, (354) Eleonora, and (4125) Lew Allen are obtained 
from \citet{2011ApJ...741...68M}. Albedo values for asteroids (289) Nenetta, (1951) Lick, and (4490) Bambery are obtained from \citet{2004PDSS...12.....T}. A slope parameter G=0.15 has been adopted for all the asteroids 
with the exception of (354) Eleonora, which has G=0.37 \citep{2011ApJ...741...68M}. If the albedo and $\eta$ of the asteroid are unknown a value of $p_{v}=0.2$ and $\eta=1$ is assumed.}}
\begin{center}\footnotesize
\hspace*{-1.9cm}
\begin{tabular}{|c|c|c|c|c|c|c|c|c|}

\hline
Object&BIC ($\mu$m)&$\Delta$BIC ($\mu$m)&Fo (mol \%)&$\Delta$Fo (mol \%)&BI$_{\rm depth}$ (\%)&$p_{v}$&$\eta$&T (K)   \\  \hline
246 Asporina&1.065&1.070&66.3&56.5&24.00&$0.2069\pm0.0294$&$1.118\pm0.018$&160.3 \\
289 Nenetta&1.069&1.074&58.5&48.8&48.08&$0.2438\pm0.042$&---&149.8 \\
354 Eleonora&1.060&1.065&76.0&66.3&27.04&$0.1732\pm0.0324$&$1.071\pm0.084$&161.2 \\
1951 Lick&1.061&1.063&74.1&70.2&38.35&$0.0895\pm0.020$&---&247.5 \\
4125 Lew Allen&1.062&1.065&72.1&66.3&23.56&$0.1462\pm0.0358$&$1.249\pm0.033$&204.3 \\
4490 Bambery&1.066&1.070&64.3&56.5&29.11&$0.2156\pm0.024$&---&204.0 \\ 
5261 Eureka&1.071&1.074&54.6&48.8&27.70&---&---&225.6 \\
 \hline

\end{tabular}\hspace*{-1.8cm}
\end{center}
\end{table}

\begin{figure*}[!ht]
\begin{center}
\includegraphics[height=10cm]{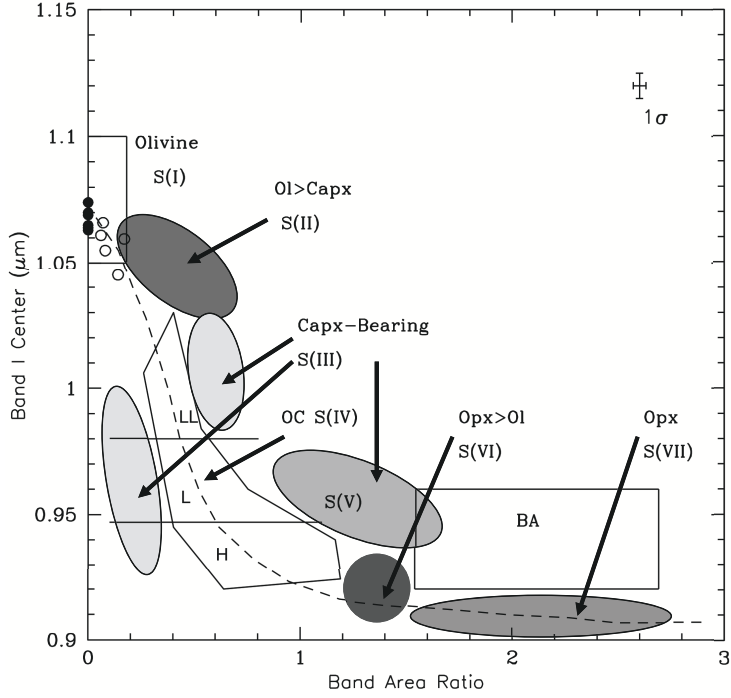}
\caption{\label{f:BICBAR3A} {\small Plot of the Band I center versus BAR for the monomineralic-olivine (filled circles) and olivine-rich (open circles) asteroids studied in the present work. The average 1-$\sigma$ error bars 
are shown in the upper right corner. The rectangular zone corresponding to the S(I)-type asteroids encompasses monomineralic olivine assemblages \citep{1993Icar..106..573G}. The rectangular region (BA) includes the 
pyroxene-dominated basaltic achondrite assemblages. The polygonal region, corresponding to the S(IV) subgroup, represents the mafic silicate components of ordinary chondrites (OC). The dashed curve indicates 
the location of the olivine-orthopyroxene mixing line \citep{1986JGR....9111641C}. The horizontal lines represent the approximate boundaries for ordinary chondrites found by \citet{2010Icar..208..789D}.}}
\end{center}
\end{figure*}

The primary diagnostic feature in the spectrum of olivine assemblages is composed of three overlapping features and is centered near $\sim$ 1 $\mu$m. This composite feature is attributed to electronic transitions of 
Fe$^{2+}$ occupying both the M1 and M2 crystallographic sites \citep{1993macf.book.....B}. Using NIR diffuse spectral reflectance, \citet{1987JGR....9211457K} investigated apparent wavelength shifts as a function of 
mineral chemistry in the Fe/Mg olivine series from forsterite Fo$_{11}$ to Fo$_{91}$. They found that the composite 1 $\mu$m feature in olivine moves to longer wavelengths as the Fe$^{2+}$ content increases.

\begin{figure*}[!ht]
\begin{center}
\includegraphics[height=10cm]{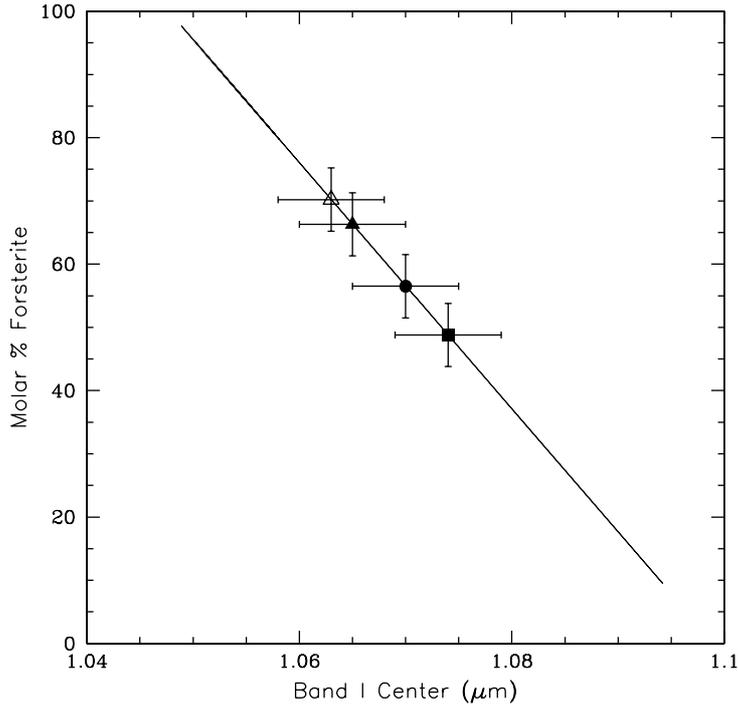}
\caption{\label{f:FobIc} {\small Improved olivine calibration plot developed by \citet{2011P&SS...59..772R}. The linear relationship between the Band I center and forsterite content is represented as a solid line. The root 
mean square error between the molar content of forsterite determined using Eq. (1) and the laboratory measurements is $5\%$. Also shown the calculated forsterite content for (246) Asporina and (4490) Bambery 
(both depicted as a filled circle), (289) Nenetta and (5261) Eureka (both depicted as a filled square), (354) Eleonora and (4125) Lew Allen (both depicted as a filled triangle), and (1951) Lick (open triangle).}}
\end{center}
\end{figure*}

\citet{2011P&SS...59..772R} improved the olivine calibration developed by \citet{1987JGR....9211457K} by adding more Mg-rich olivine (Fo$_{85-93}$) samples. Reflectance spectra of these additional samples were 
obtained at the University of Winnipeg HOSERLab. This calibration plot is shown in Figure \ref{f:FobIc}, where the solid line represents a linear fit to the data used by \citet{2011P&SS...59..772R}. The equation that 
describes the linear fit is given by

\begin{equation}
Fo = -1946.6\times (BIC)+2139.4
\end{equation}
 
Using this equation, along with the measured Band I centers (BIC), we estimated the molar \% of forsterite (Fo) for the monomineralic-olivine asteroids. The root mean square error between the molar content of forsterite 
determined using Eq. (1) and the laboratory measurements is $5\%$, however due to the band center uncertainty the percentage of forsterite can only be estimated with a precision of $\pm$10-15\%.

The laboratory measurements used to develop Eq. (1) were obtained at room temperature (300 K), however the surface temperature of asteroids is typically much lower. The effects on spectral parameters of mafic minerals 
due to temperature have been investigated by several authors \citep[e.g.,][]{1985JGR....9012434S,1999AdSpR..23.1253S,2000Icar..147...79M,2002Icar..155..169H, 2012Icar..217..153R, 2012Icar..220..36S}. These effects 
are seen as broadening or narrowing of the absorption features and shifting of the band centers. In order to determine whether this temperature difference could affect the Band I centers measured from asteroid spectra, we 
have reanalyzed spectra of olivine (Fo$_{86}$) from \citet {2002Icar..155..169H} acquired in the temperature range between 80 and 400 K, with a temperature resolution of 20 K. Band I centers were measured using the 
same procedure applied to the asteroid spectra and then plotted as a function of temperature. We found that the Band I center shifts to longer wavelengths as the temperature increases. This shift in Band I center can be 
described as 

\begin{equation}
BIC(\mu m)=(1.18\times 10^{-7})T^{2}-(2.15\times 10^{-5})T+1.05
\end{equation}

From this equation we derived a wavelength correction with respect to room temperature (300 K) for the Band I center

\begin{equation}
\Delta BIC(\mu m)=-(1.18\times 10^{-7})T^{2}+(2.15\times 10^{-5})T+0.004
\end{equation}

This correction must be added to the calculated Band I center of each asteroid. The average surface temperature of the asteroids was calculated in the same way as \citet{2009M&PS...44.1331B}. Thus, after correcting 
the Band I centers, Eq. (1) was used to calculate the olivine composition of the asteroids. The temperature corrected Band I centers, olivine compositions (with and without temperature corrections) and the 
average surface temperature of the asteroids are presented in Table \ref{t:Table2}. The temperature corrected Band I centers of the monomineralic-olivine asteroids are depicted in Fig. \ref{f:BICBAR3A} as filled circles.

\subsubsection{Mineralogical analysis}

\citet{1984Sci...223..281C} were the first to obtain spectrophotometric observations in the NIR of (246) Asporina using the NASA IRTF, which led to the identification of nearly monomineralic-olivine on an asteroid surface. 
Using spectral curve matching techniques, the authors suggested that Asporina's spectrum best matched lab samples of olivine that were coarse grains mixed with "substantial fraction of grains $<$ 45 $\mu$m in 
size" \citep{1984Sci...223..281C}. Their best match for (246) Asporina was an olivine assemblage with mineralogy of Fo$_{60-90}$. More recently, \citet{2007M&PS...42..155S} estimated the composition 
of (246) Asporina to be Fo$_{63}$. The spectrum of (246) Asporina shown in Figure \ref{f:Aspec1} is dominated by a deep complex feature (band depth 24\%) at $\sim$ 1 $\mu$m, and the derived temperature-corrected 
 band center for this feature is 1.07 $\mu$m. The presence of this complex 1 $\mu$m feature, composed of three overlapping narrower features, is a strong indication of the presence of 
 olivine \citep{1993macf.book.....B} with no high- (Type B CPX) or low-Ca pyroxene due to the lack of 2 $\mu$m feature. Using the calculated Band I center, and based on the olivine calibration by 
 \citet{2011P&SS...59..772R} (Eq. 1) we found that  Asporina has an estimated forsterite abundance of Fo$_{56.5\pm5}$. This value, depicted in Fig. \ref{f:FobIc} as a filled circle, is a little lower than the forsterite value 
 reported for Asporina (Fo$_{63}$) by \citet{2007M&PS...42..155S}. 

\begin{figure*}[!ht]
\begin{center}
\includegraphics[height=10cm]{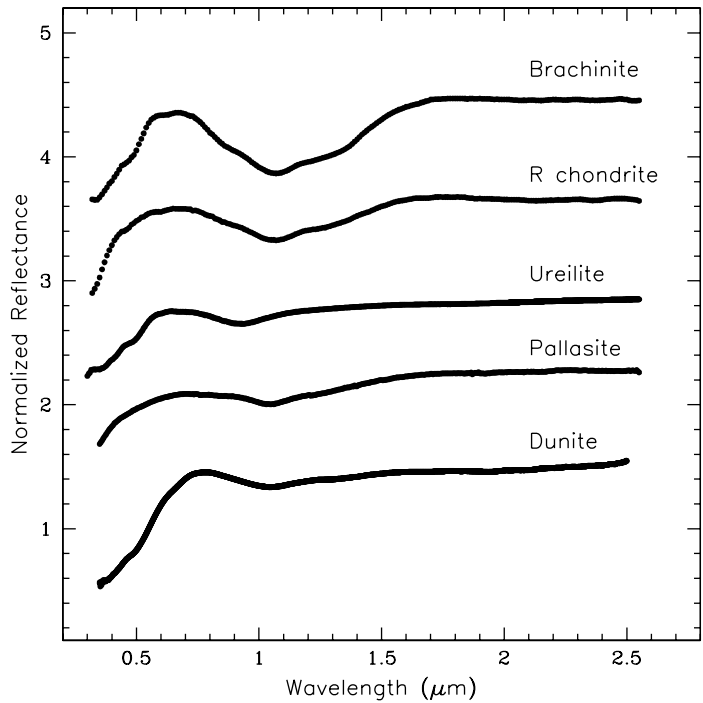}
\caption{\label{f:metspec3} {\small VIS-NIR spectra of olivine-dominated meteorites obtained from RELAB. All spectra are normalized to unity at 1.5 $\mu$m and have been offset for clarity. From the bottom to the 
top: Dunite NWA 2968 from Vesta (measured at the University of Winnipeg HOSERLab), pallasite (Marjalahti, RELAB ID:MS-CMP-005-X), ureilite (Y-791538, RELAB ID:MP-TXH-101), R chondrite (Rumuruti, RELAB 
ID:MT-TJM-013), and Brachinite (EET99402, RELAB ID:TB-TJM-058).}}
\end{center}
\end{figure*}

After determining the olivine composition of Asporina the next step is to identify a possible meteorite analogue. S(I)-type asteroids have been traditionally interpreted as pieces of asteroid mantles based on the 1 $\mu$m 
olivine spectral feature. While pallasites are often invoked as possible analogues, they are typically equated to core-mantle boundaries, and would therefore represent very deep mantle samples. Other olivine-rich 
meteorites that could be potential analogues for S(I) asteroids include the R-chondrites, brachinites, and ureilites. Representative spectra of these meteorites are shown in Fig. \ref{f:metspec3}, for comparison we also 
included the spectrum of dunite NWA 2968 from Vesta measured at the University of Winnipeg HOSERLab for this study. This meteorite shows FeO/MnO ratios and oxygen-isotopic compositions consistent with howardites, 
eucrites and diogenites (HED) meteorites, and therefore it is considered to be a fragment of Vesta's mantle.

Pallasites consist of metal and silicate (olivine) in roughly equal amounts with troilite as a minor phase \citep{1998PlanetaryMaterialsM}. The three types of pallasites (main group, Eagle station, and pyroxene-pallasite) 
differ from each other based on mineralogy, composition, and O-isotopes of the silicate and metal components. All three pallasite groups have olivine with Fo$_{80-90}$ 
\citep{1998PlanetaryMaterialsM, 1997M&PS...32..231R}, which is much higher than Asporina. Work by \citet{1990JGR....95.8323C} has shown that silicate spectral features are severely suppressed with the addition of 
metal, and significant metal ($>$50\%) can cause reddening of spectral slope. An example of a pallasite spectrum is depicted in Fig. \ref{f:metspec3} that shows the VIS-NIR spectrum of pallasite Marjalahti obtained from 
the RELAB database \citep{2004LPI....35.1720P}. Based on the mismatch in olivine chemistry, and spectral parameters (band depth suppression and red spectral slope) we conclude that pallasites are not good meteorite 
analogues for (246) Asporina. 

Ureilites are composed mostly of mafic silicates like olivine (60-90\%) and minor pyroxene which are embedded in a dark matrix of carbonaceous material (graphite) \citep{1998PlanetaryMaterialsM}. Olivine is 
Mg-rich (Fo$_{74-95}$) and pyroxene is in equilibrium \citep{1998PlanetaryMaterialsM}. Spectra of most ureilites exhibit absorption features near 1 and 2 $\mu$m due to olivine and pyroxene, but because of the 
presence of carbon these features are severely suppressed (Band I depth 6-12\%, Band II depth 1-3\%) (see Fig. \ref{f:metspec3}). Most ureilites have moderate albedos between 0.10-0.14, and spectrally they 
exhibit flat or negative slopes across the 0.7-2.6 {$\mu$}m interval \citep{2010M&PS...45.1668C}. In contrast, 246 Asporina has an olivine chemistry of Fo$_{56.5\pm5}$, an albedo of $\sim$ 0.21, and a Band I depth of 
24\%, which are inconsistent with ureilites. Based on the differences in chemistry and band parameters we conclude that ureilites are not analogues for 246 Asporina. 


R chondrites (Fo$_{60-63}$) \citep{1998PlanetaryMaterialsB} and brachinites (Fo$_{60-75}$) (Mittlefehldt et al., 1998) are olivine-dominated meteorites with chemistries within the range of the Asporina's 
estimated chemistry. R chondrites are formed under highly oxidizing nebular processes with most olivine having a chemistry of Fo$_{60-63}$ with minor clinopyroxene ($\sim$5\%) and traces of low-Ca pyroxene 
\citep{1994Metic..29..275S}. Brachinites are formed in igneous processes and contain 80-90\% olivine with 5-10\% clinopyroxene and traces of low-Ca pyroxene \citep{2003M&PS...38.1601M}. Examples of R chondrite 
and brachinite spectra are depicted in Fig. \ref{f:metspec3}. Based on Asporina's olivine chemistry Fo$_{56.5\pm5}$, and the lack of a significant 2 $\mu$m feature due to pyroxene, brachinites are the most probable 
meteorite analogues for this asteroid, although R chondrites can not be completely ruled out. The moderate albedo of 0.21 for (246) Asporina also supports this linkage with brachinites, which have relatively moderate 
albedos of $\sim$0.19. The albedo of brachinites was estimated from the lab spectra of a $<$150 $\mu$m size powder of Brachina, obtained at the University of Winnipeg HOSERLab for this study, and EET99402 from 
the RELAB database. \citet{2007M&PS...42..155S} modeled the formation of brachinites from an R chondrite precursor using MELTS model \citep{1998Asimow, 1995CoMP..119..197G} and concluded that an olivine-rich 
residue with Fa concentrations similar to brachinites can be produced at $\sim$25\% melting at a temperature of $\sim$1220 $^\circ$C. This suggests that Fe-rich olivines like those in brachinites can be created starting 
with an R chondrite type material.

For asteroid (289) Nenetta, the temperature corrected Band I center is located at 1.074 $\mu$m with a Band I depth of 48.08\%. Using Eq. (1) we determined that the olivine composition of this asteroid is Fo$_{48.8\pm5}$. 
This composition, depicted in Fig. \ref{f:FobIc} as a filled square, is consistent with the range reported by \citet{1984Sci...223..281C} (Fo$_{40-80}$), and similar to that found by \citet{2012DPS....4411011L} (Fo$_{53}$), 
however it is lower than the value obtained by \citet{2007M&PS...42..155S} (Fo$_{60}$). Based on its olivine chemistry, the lack of a significant 2 $\mu$m feature, and its albedo of 0.24, brachinites are the most likely 
meteorite analogues. However, as in the case of (246) Asporina, R chondrites can not be completely ruled out. It is also important to point out that the olivine composition of (289) Nenetta seems to be more ferroan than the 
typical compositions observed among brachinites and R chondrites.   

The spectrum of asteroid (354) Eleonora has a Band I center at 1.065 $\mu$m and Band I depth of 27.04\%. For this object we obtained an olivine composition of Fo$_{66.3\pm5}$ (Fig. \ref{f:FobIc}, filled triangle). 
\citet{2007M&PS...42..155S} found that this asteroid is a magnesian object, however their estimated composition of Fo$_{92}$ is much higher than ours. The composition of (354) Eleonora determined by us seems to be 
consistent with the range found for brachinites (Fo$_{60-75}$). 

Asteroid (1951) Lick is the only near-Earth asteroid (NEA) among the studied objects. The Band I center of (1951) Lick is located at 1.063 $\mu$m (after temperature correction) and its Band I depth has a value 
of 38.35 \%. Using MGM, \citet{2007M&PS...42..155S} found that the composition of this asteroid is consistent with magnesian olivine (Fo$_{79}$). \citet{2007A&A...472..653B} determined that the best fit of Lick's 
spectrum is obtained by a linear combination of reflectance spectra of San Carlos olivine (20\%) and meteorite Brachina (80\%), along with a fraction of nanophase iron needed to account for space weathering 
effects. Our analysis of (1951) Lick using Eq.(1) yielded a composition of Fo$_{70.2\pm5}$, which is the highest forsterite value found among the monomineralic-olivine asteroids studied in this work. The olivine chemistry of 
(1951) Lick, represented in Fig. \ref{f:FobIc} as an open triangle, is consistent with the composition of brachinites.  

In the case of asteroid (4125) Lew Allen, the Band I center is located at 1.065 $\mu$m and the Band I depth is 23.56\%. According to our calculations, the olivine chemistry for this asteroid is Fo$_{66.3\pm5}$ 
(Fig. \ref{f:FobIc}, filled triangle). No previous work was performed on this object. Based on its olivine chemistry and the absence of a significant 2 $\mu$m feature, brachinites are the best meteorite analogues for this 
asteroid.  

For asteroid (4490) Bambery, the Band I center is located at 1.070 $\mu$m with a Band I depth of 29.11\%. As in the previous cases, the olivine composition of the asteroid was calculated, given the value 
of Fo$_{56.5\pm5}$ (Fig. \ref{f:FobIc}, filled circle). Like Lew Allen, this is the first time that the composition of this asteroid is determined. Brachinites and R chondrites have olivine compositions that, within the uncertainty, 
are consistent with the value found for Bambery. However, considering the lack of the 2 $\mu$m feature and its albedo of $\sim$ 0.22, brachinites are the most probable meteorite analogues for this object.

Asteroid (5261) Eureka is the largest known Mars Trojan \citep{2007Icar..192..442T}. \citet{2007Icar..192..434R} obtained NIR spectra of Eureka with the IRTF and SpeX. From their analysis they identified angrites 
as possible meteorite analogues for this asteroid. Angrites are a rare group of basaltic achondrites with mineralogies dominated by CaO-rich olivine, Ca-Al-Ti-rich pyroxene and 
anorthite \citep{2006mess.book...19W}. Spectrally, they exhibit a broad absorption feature centered around 1 $\mu$m and, in some cases, a weak absorption feature near 2 $\mu$m \citep{2006M&PS...41.1139B}. Using 
 Hapke formalism, \citet{2007Icar..192..434R} modeled the reflectance spectrum of Eureka by combining angrites and neutral components. Although they obtained a reasonably good fit for Eureka's spectrum they couldn't 
 rule out R chondrites as possible meteorite analogues. The thermal-IR spectrum (5-30 $\mu$m) of Eureka obtained with the Spitzer IRS by \citet{2011epsc.conf.1199L} showed olivine reststrahlen features similar 
 to those observed from laboratory spectra of R chondrites, brachinites and chassignites. From their study \citet{2011epsc.conf.1199L} determined that the molar content of Fo for Eureka is probably not higher than 
 Fo$_{65}$. The measured Band I center for Eureka is located at 1.074 $\mu$m with a band depth of 27.7\%. Using Eq. (1) we calculated its olivine composition, giving a value of Fo$_{48.8\pm5}$ (Fig. \ref{f:FobIc}, filled 
 square). Similar to asteroid (289) Nenetta, the olivine composition of Eureka is more ferroan than the typical compositions found among olivine-rich meteorites. However, based on these results and the lack of a 
 significant 2 $\mu$m feature, brachinites are the most likely meteorite analogues for Eureka, although R chondrites can not be completely ruled out.     
 
In general, olivine compositions of asteroids obtained using Eq. (1) seems to be more Fe-rich than those estimated by previous work using MGM \citep[e.g.,][]{1998JGR...10313675S, 2007M&PS...42..155S}. 
These differences regarding the estimation of forsterite chemistry based on the Band I center by \citep[e.g.,][]{1998JGR...10313675S, 2007M&PS...42..155S} and King and Ridley (1987) are due to the way the band 
center is calculated. The band centers calculated using MGM are at shorter wavelengths than those of \citet{1987JGR....9211457K}. In addition, the temperature correction that we have applied to the asteroid Band I 
centers using Eq. (3) slightly increases their value, hence the molar content of Fo calculated using Eq. (1) will decrease. 

As noted earlier, monomineralic olivine can be formed through different mechanism: accretion of grains from an oxidized nebular region without significant post-accretionary heating (R chondrites), partial melting with 
extraction of a basaltic melt leaving an olivine-rich residue (brachinites), or crystallization of olivine from a melt to form an olivine mantle or olivine-rich layer (differentiated object). The Fe content of igneously formed olivine 
depends on the Fe content of the starting parent material with the Fe content of the final olivine never being higher than the precursor material. For example, olivine produced in an igneous process with an LL-chondrite 
precursor material can never have Fe content more than Fo$_{72}$. In other words, the only way to produce a Fe-rich olivine via igneous processing is to start with a precursor material that has high Fe 
content \citep{2007M&PS...42..155S}. Based on the results obtained, and taking into account the uncertainties associated to the Band I centers and Fo content, it is possible that at least three of the studied asteroids (354 
Eleonora, 1951 Lick and 4125 Lew Allen) could have formed from melting of ordinary chondrite-like bodies.


\subsection{Olivine-rich asteroids}

Asteroids that have been grouped in this category are: (446) Aeternitas, (863) Benkoela, (984) Gretia, (2501) Lohja, and (3819) Robinson. In addition to the 1 $\mu$m feature, the spectra of these objects exhibit a weak 
absorption feature at $\sim$ 2 $\mu$m indicative of the presence of a second mineral in the assemblages, most likely pyroxene. For each VIS-NIR spectrum spectral band parameters, band centers, Band Area Ratios 
(BAR), and band depths along with their errors were measured in the same way as in \citet{2012Icar..220..36S, 2013Icar..225..131S}. The band parameters with their corresponding errors are presented in 
Table \ref{t:Table3}. The Band I centers and BAR of the olivine-rich asteroids are depicted in Fig. \ref{f:BICBAR3A} as open circles.

\begin{table}[!ht]
\caption{\label{t:Table3} {\small Spectral band parameters for the olivine-rich asteroids. The columns in this table correspond to: object number and designation, 
Band I center (BIC$\pm$0.005), Band I depth (BI$_{\rm{depth}}\pm$0.3), Band II center (BIIC$\pm$0.01), Band II depth (BII$_{\rm{depth}}\pm$0.5), Band Area 
Ratio (BAR$\pm$0.03), olivine-pyroxene abundance ratio (ol/(ol+px)$\pm$0.003), the ratio of low-Ca pyroxene to total pyroxene (LCP/(LCP+HCP)$\pm$0.09), and geometric albedo ($p_{v}$). 
In the case of the olivine-rich asteroids no temperature corrections to the band parameters were applied because spectra of R chondrites obtained at different temperatures are not available. Albedo 
values are obtained from \citet{2011ApJ...741...68M}.}}
\begin{center}\footnotesize
\hspace*{-1.9cm}
\begin{tabular}{|c|c|c|c|c|c|c|c|c|}

\hline

Object&BIC ($\mu$m)&BI$_{\rm depth}$ (\%)&BIIC ($\mu$m)&BII$_{\rm depth}$ (\%)&BAR&ol/(ol+px)&LCP/(LCP+HCP)&$p_{v}$ \\  \hline
446 Aeternitas&1.066&38.00&1.980&3.00&0.07&0.92&0.90&$0.1902\pm0.0492$ \\
863 Benkoela&1.059&33.50&1.982&6.50&0.17&0.68&0.89&$0.1123\pm0.0163$ \\
984 Gretia&1.045&31.64&2.040&4.77&0.14&0.78&0.60&$0.3990\pm0.0930$ \\
2501 Lohja&1.061&41.90&2.022&3.20&0.06&0.93&0.69&$0.1898\pm0.0440$\\
3819 Robinson&1.055&39.20&2.026&4.35&0.08&0.91&0.67&$0.3580\pm0.2982$\\
 \hline

\end{tabular}\hspace*{-1.8cm}
\end{center}
\end{table}

\begin{table}[!ht]
\caption{\label{t:Table4} {\small Spectral band parameters for the R chondrites. The columns in this table correspond to: meteorite name, RELAB ID, 
Band I center (BIC$\pm$0.003), Band II center (BIIC$\pm$0.005), Band Area Ratio (BAR$\pm$0.01), olivine-pyroxene abundance ratio ol/(ol+px), and the 
ratio of low-Ca pyroxene to total pyroxene LCP/(LCP+HCP), both derived from laboratory measurements.}}
\begin{center}\small
\begin{tabular}{|c|c|c|c|c|c|c|}

\hline

Meteorite&RELAB ID&BIC ($\mu$m)&BIIC ($\mu$m)&BAR&ol/(ol+px)&LCP/(LCP+HCP) \\  \hline
NWA753&TB-TJM-114&1.061&1.951&0.08&---&--- \\
LAP04840&DD-AHT-107&1.032&1.954&0.01&0.96$^{a}$&1.00$^{a}$ \\
Rumuruti&MT-TJM-013&1.067&2.121&0.06&0.93$^{b}$&0.09$^{b}$ \\
PRE95411&MT-TXH-045&1.066&1.978&0.20&---&--- \\
PCA91002&MB-TXH-065-A&1.065&2.072&0.13&0.81$^{c}$&0.57$^{c}$ \\
ALH85151&MB-TXH-045&1.068&2.060&0.17&0.68$^{d}$&0.53$^{d}$ \\
A-881988&MP-TXH-059&1.051&1.968&0.03&---&--- \\

 \hline

\end{tabular}
\end{center}

{\small $^{a}$ Data from \citet{2008GeCoA..72.5757M}}

{\small $^{b}$ Data from \citet{1994Metic..29..275S}}

{\small $^{c}$ Data from \citet{1994Metic..29..255R}}

{\small $^{d}$ Data from \citet{1996GeCoA..60.2243K}}

\end{table}

\begin{figure*}[!ht]
\begin{center}
\includegraphics[height=10cm]{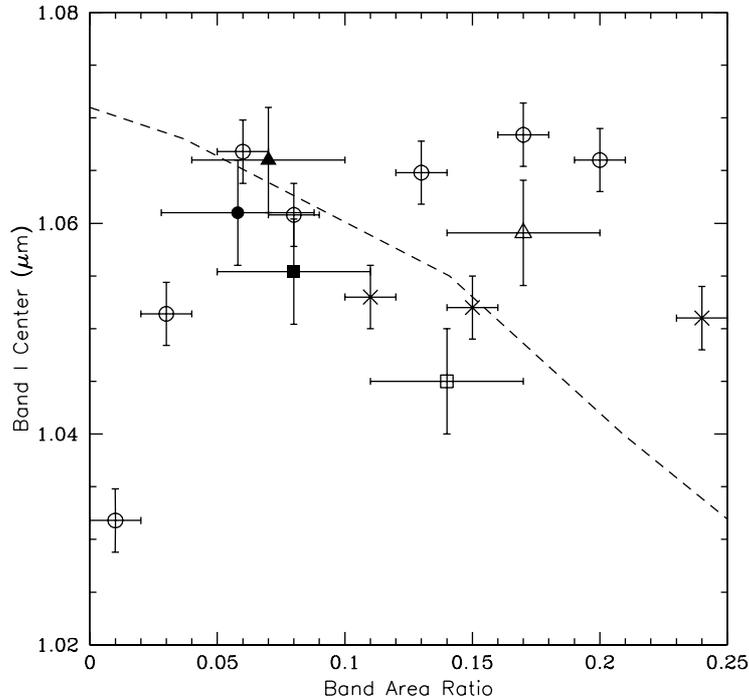}
\caption{\label{f:bIbar} {\small Band I center vs. Band Area Ratio (BAR) for olivine-rich asteroids and R chondrites. Values measured for R chondrites are depicted with open circles. The asteroids whose band 
parameters have been plotted are: (2501) Lohja (filled circle),  (3819) Robinson (filled square), (446) Aeternitas (filled triangle), 863 Benkoela (open triangle) and (984) Gretia (open square). For comparison we have 
included measured values (x symbols) from a mixture of OLV003 (forsterite)+PYX016 (diopside) obtained at the University of Winnipeg HOSERLab. The dashed curve indicates the location of the olivine-orthopyroxene 
mixing line from \citet{1986JGR....9111641C}.}}
\end{center}
\end{figure*}

\begin{figure*}[!ht]
\begin{center}
\includegraphics[height=10cm]{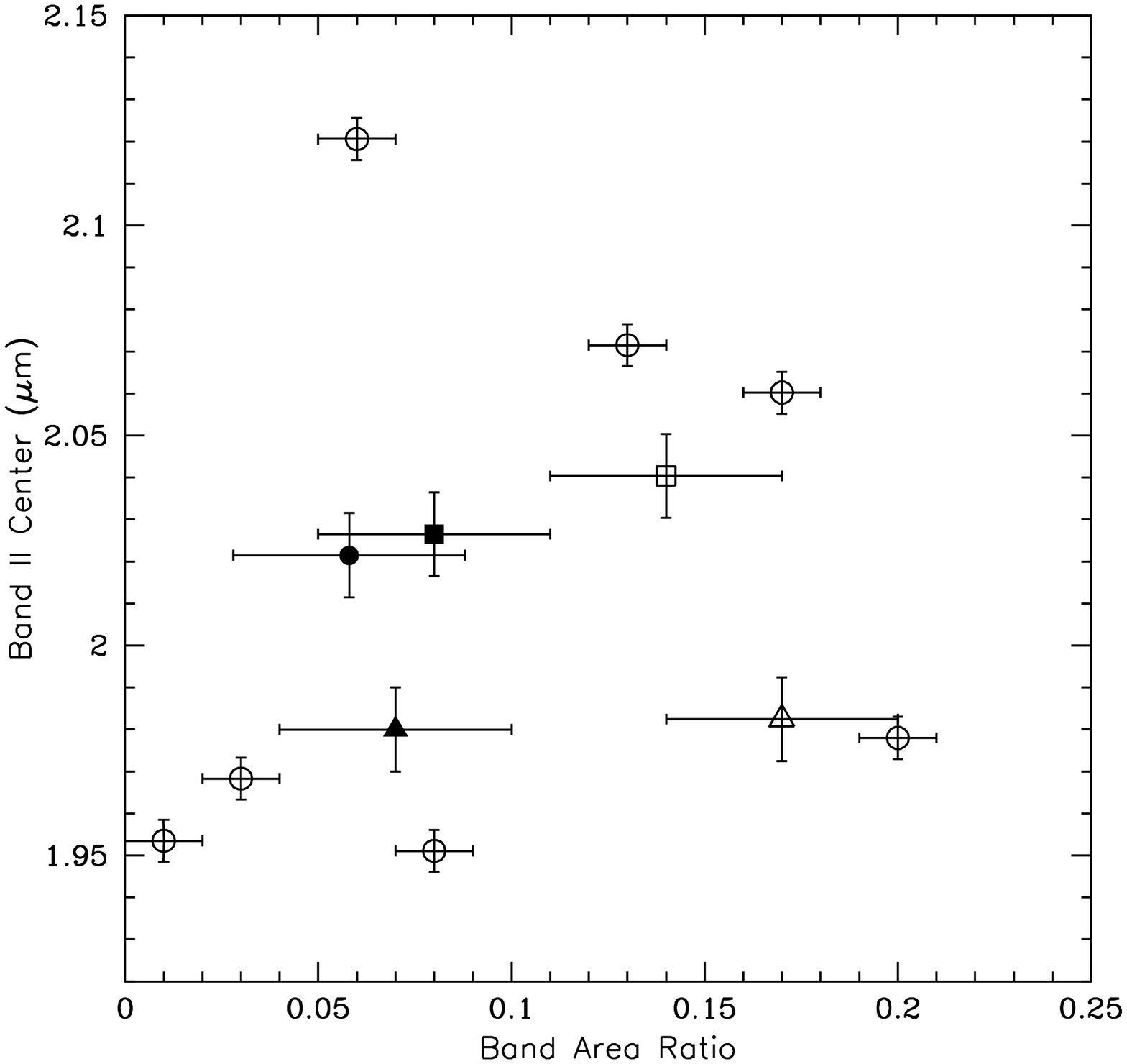}
\caption{\label{f:bIIbar} {\small Band II center vs. Band Area Ratio (BAR) for olivine-rich asteroids and R chondrites. Values measured for R chondrites are depicted with open circles. The asteroids whose band 
parameters have been plotted are: (2501) Lohja (filled circle),  (3819) Robinson (filled square), (446) Aeternitas (filled triangle), 863 Benkoela (open triangle) and (984) Gretia (open square).}}
\end{center}
\end{figure*}

The most common secondary or accessory mineral present in olivine-rich asteroids is low-Ca pyroxene, which gives rise to two absorption features at $\sim$0.9 and 2 $\mu$m. Depending on the type of pyroxene and its 
Fe content, the band center of the 1 $\mu$m olivine feature is shifted to shorter wavelengths. From the calibration developed by \citet{2011P&SS...59..772R}, the band center of the 1 $\mu$m olivine feature ranges from 
$\sim$1.048 $\mu$m for pure forsterite to $\sim$ 1.1 $\mu$m for pure fayalite. Depending on the Fo content of the olivine and the abundance of low-Ca pyroxene in a olivine + low-Ca pyroxene assemblage, the Band I 
center could be shifted beyond this band center range ($\sim$1.048-1.10 $\mu$m) leading to inconsistent Fo estimates using this calibration. Even if the measured Band I centers are within this range the presence of 
pyroxene might shift the Band I center to longer or shorter wavelength (depending on the type of pyroxene and the cation abundance) preventing an accurate determination of olivine chemistry using Eq. (1). Therefore, we 
decided not to use this calibration with the olivine-rich asteroids and to look for possible meteorite analogs that could be used to develop a new calibration.

Olivine-rich meteorites that could be potential analogs for these asteroids include pyroxene pallasites, brachinites, ureilites, and R-chondrites. Pyroxene pallasites are dominated by olivine (55-63 vol \%) and contain 
(1-3 vol \%) low-Ca pyroxenes \citep{1998PlanetaryMaterialsM}. Due to the presence of metal, sample preparation for laboratory spectroscopy is extremely difficult for metal-silicate mixtures like 
pallasites \citep{1976JGR....81..905G, 1990JGR....95.8323C}. As a result there are no useful NIR laboratory spectra of both metal and olivine+pyroxene from a pyroxene pallasite. Due to this limitation, these meteorites are 
not considered in the present analysis. As it was mentioned earlier, brachinites contain $\sim$ 5-10\% clinopyroxene and traces of low-Ca pyroxene, however brachinite spectra don't show the 2 $\mu$m feature. Spectra of 
most ureilites show absorption features near 1 and 2 $\mu$m, however these features are severely suppressed. Band depths measured for the olivine-rich asteroids in our sample range from $\sim$ 
32-42\% (Band I depth), and from $\sim$ 3-7\% (Band II depth). Furthermore, the typical albedos for ureilites (0.10-0.14) are lower than most of olivine-rich asteroids. R chondrites, on the other 
hand, have an olivine abundance of typically 65-78 vol \% and different proportions of low-Ca and Ca-pyroxene \citep{2011ChEG...71..101B}. Spectrally they show the 
1 $\mu$m feature and a weak absorption feature at $\sim$ 2 $\mu$m. In Figs. \ref{f:bIbar} and \ref{f:bIIbar} we plotted the Band I and Band II centers vs. BAR, respectively of seven R chondrite spectra from the RELAB 
database. Measured values are depicted as open circles and have been plotted along with the values measured from the olivine-rich asteroid spectra. Spectral band parameters of the R chondrites are presented in Table 
\ref{t:Table4}. An inspection of Figs. \ref{f:bIbar} and \ref{f:bIIbar} shows that the band centers and BAR values of the olivine-rich asteroids are located within the range of those measured for the R chondrites. As can 
be seen in Fig. \ref{f:bIbar}, values obtained from the R chondrites deviate from the olivine-orthopyroxene mixing line of \citet{1986JGR....9111641C} (dashed curve). This deviation could be explained by the 
presence of a clinopyroxene component in the R chondrites. This behavior is also observed in the measured values of an olivine-clinopyroxene mixture depicted as "x" symbols in Fig. \ref{f:bIbar}. These values 
correspond to a mixture in which the fraction of olivine to clinopyroxene ranges from 80:20 to 40:60. Based on these results we developed a new set of calibrations in an effort to determine the olivine abundance and the 
ratio of low-Ca pyroxene to total pyroxene of the olivine-rich asteroids.

\begin{figure*}[!ht]
\begin{center}
\includegraphics[height=10cm]{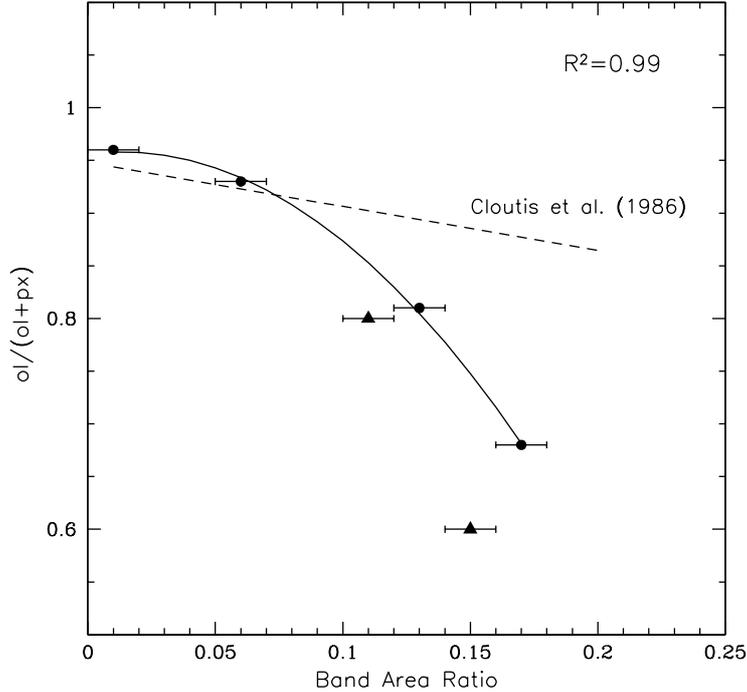}
\caption{\label{f:olbar} {\small ol/(ol+px) ratios vs. Band Area Ratio (BAR) measured for R chondrites (filled circles). The solid line represents a second order polynomial fitted to the data. The coefficient of determination 
(R$^{2}$) is given. The root mean square error between the ol/(ol+px) ratios determined using Eq. (4) and the laboratory measurements is 0.003. For comparison we have included the linear relationship between 
ol/(ol+opx) and BAR from \citet{1986JGR....9111641C} (dashed line) given by $ol/(ol+opx)=-0.417\times BAR+0.948$. Also shown, as filled triangles, measured values from a mixture of OLV003 (forsterite)+PYX016
(diopside) obtained at the University of Winnipeg HOSERLab.}}
\end{center}
\end{figure*}

Figure \ref{f:olbar} shows the ol/(ol+px) ratios vs. BAR measured for R chondrites. Here the ol/(ol+px) values correspond to the abundances of olivine and LCP and HCP. Modal 
abundances (included in Table \ref{t:Table4}) are only available for four samples, however from Fig. \ref{f:olbar} it is possible to see how the ol/(ol+px) ratio decreases as the BAR increases. This curve (solid line) 
shows a much steeper decrease in olivine content with increasing BAR than shown by the \citet{1986JGR....9111641C} olivine pyroxene mixing experiments (dashed line). This discrepancy could be attributed to the 
fact that the linear relationship obtained by \citet{1986JGR....9111641C} was derived from an olivine-orthopyroxene mixture, while our measurements include both LCP and HCP. As an example we have included in 
Fig. \ref{f:olbar} measured values from an olivine-clinopyroxene mixture (depicted as filled triangles), which show a pronounced deviation from the linear relationship obtained by \citet{1986JGR....9111641C}. The 
correlation between the ol/(ol+px) ratio and the BAR found for the R chondrites can be described by a second order polynomial fit:

\begin{equation}
ol/(ol+px)=-11.27\times BAR^{2}+0.302\times BAR+0.956
\end{equation}

With this equation we determined the olivine abundance of the olivine-rich asteroids from their BAR values. Calculated ol/(ol+px) ratios are presented in Table \ref{t:Table3}. 

\begin{figure*}[!ht]
\begin{center}
\includegraphics[height=10cm]{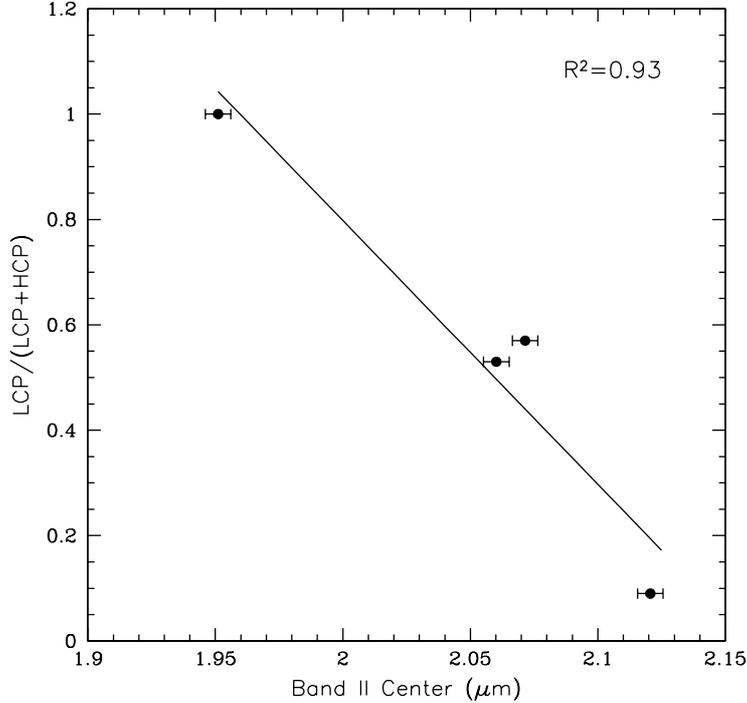}
\caption{\label{f:opxbIIc} {\small LCP/(LCP+HCP) ratios vs. Band II center measured for R chondrites. The solid line represents a linear fit to the data. The coefficient of determination (R$^{2}$) is given. The root mean 
square error between the (LCP/(LCP+HCP)) ratios determined using Eq. (5) and the laboratory measurements is 0.09.}}
\end{center}
\end{figure*}

In Figure \ref{f:opxbIIc} we plotted LCP/(LCP+HCP) ratios vs. Band II centers measured for R chondrites. Here the LCP/(LCP+HCP) values correspond to the ratio of low-Ca pyroxene to total pyroxene. As can be 
seen in Fig. \ref{f:opxbIIc}, Band II centers move to longer wavelengths with increasing Ca$^{2+}$ content. This shift in the wavelength position is explained by crystal field theory (CFT), and is due to the effects of replacing 
the larger Ca$^{2+}$ cation into the crystallographic site of the smaller Fe$^{2+}$ cation, resulting in changes in the crystal structure and thus in the crystal fields and electronic transitions which produce this
absorption feature \citep{1993macf.book.....B}. A least-squares fit of the data in Fig. \ref{f:opxbIIc} yields
 
\begin{equation}
LCP/(LCP+HCP)=-5.006\times(BIIC)+10.81
\end{equation}

Using equation (5) we determined the fraction of low-Ca pyroxene present in the olivine-rich asteroids from their measured Band II centers. The LCP/ (LCP+HCP) ratios are presented in Table \ref{t:Table3}. In the case 
of the olivine-rich asteroids no temperature corrections were derived since spectra of R chondrites obtained at different temperatures are not available. It is important to stress that the results obtained using Eqs. (4) and (5) 
must be taken with caution. Although R chondrites provide a good spectral match for olivine-rich asteroids, their compositions might differ. In addition, these equations were derived from a very limited sample. Thus, these 
calculations represent only a first attempt to determine mineral abundances for this type of asteroids. Additional laboratory measurements of meteorite samples will be required to improve these spectral calibrations.


\subsubsection{Mineralogical analysis}

Of all the S(I)-type asteroids, (446) Aeternitas is one of the best-characterized members.  Based on spectral analysis, \citet{1990JGR....95.8323C} suggested that the surface assemblage 
of (446) Aeternitas contained roughly 35 wt\% metal, 55 wt\% olivine (Fo$_{80\pm10}$), 7 wt\% clinopyroxene, and 3 wt\% orthopyroxene. Similar to (246) Asporina \citep{1984Sci...223..281C}, 
\citet{1990JGR....95.8323C} suggested a substantial fine-grained ($<$45 $\mu$m) component on the surface and concluded that the best match for the spectrum of this asteroid is fine-grained olivine scattered on a 
roughened metal surface. \citet{2001M&PS...36.1587H} modeled the reflectance spectrum of Aeternitas using model mixtures of minerals that were irradiated with a pulse-laser that simulates the space 
weathering effect. They found that the best fit for Aeternita's spectrum was obtained combining 2\% fresh olivine, 93\% space-weathered olivine, 1\% space-weathered orthopyroxene, and 4\% chromite. 
\citet{2007M&PS...42..155S} suggested that (446) Aeternitas is magnesian with $\sim$5-10\% pyroxene based on the presence of a weak 2 $\mu$m feature. The VIS-NIR spectrum of 
Aeternitas exhibits a broad, asymmetric feature (Band I depth 38\%) indicating the presence of olivine as the primary mineral on the surface. The calculated Band I center is 1.066 $\mu$m. There is also a 
weak $\sim$3\% feature centered at 1.98 $\mu$m and the BAR is 0.07. From its BAR value and using Eq. (4) we determined that the ol/(ol+px) ratio for Aeternitas is 0.92. From the measured Band II center we estimated a 
LCP/(LCP+HCP) ratio value of 0.90.    

The spectrum of asteroid (863) Benkoela exhibits two absorption features centered at 1.059 $\mu$m (Band I depth 33.5\%) and 1.982 $\mu$m (Band II depth 6.5\%), with a BAR value of 0.17. For this asteroid we 
determined an olivine abundance of 0.68, the lowest among the olivine-rich asteroids studied, and a ratio of low-Ca pyroxene to total pyroxene of 0.89. 

For (984) Gretia, the Band I center is located at 1.045 $\mu$m and the Band II center is at 2.040 $\mu$m with a BAR of 0.14. Band I and II depths are 31.64\% and 4.77\%, respectively. Using Eqs. (4) and (5) we found 
that the ol/(ol+px) ratio for this asteroid is 0.78 and the LCP/(LCP+HCP) ratio is 0.60, implying that Gretia has the lowest fraction low-Ca pyroxene among the studied asteroids.

The measured Band I and Band II centers of (2501) Lohja are located at 1.061 $\mu$m (Band I depth 41.9\%)  and 2.022 $\mu$m (Band II depth 3.2\%), respectively. The calculated BAR is 0.06. With an ol/(ol+px) ratio 
of 0.93, Lohja has the highest olivine abundance among the olivine-rich asteroids studied. The fraction of low-Ca pyroxene was estimated to be 0.69. 

Like the other olivine-rich asteroids, the spectrum of (3819) Robinson exhibits two absorption features, one centered at 1.055 $\mu$m with a Band I depth of 39.20\%, and the other centered at 2.026 $\mu$m with a 
Band II depth of 4.35\%. The calculated BAR is 0.08. From its BAR value and using Eq. (4) we found that the ol/(ol+px) ratio for Robinson is 0.91. The ratio of low-Ca pyroxene to total pyroxene gave a value of 0.67.

Based on spectral matching \citet{2007M&PS...42..155S} suggested that the olivine composition of these olivine-rich asteroids is more likely to be MgO-rich. According to them this magnesian composition along with the 
position of the pyroxene feature ($<$ 2.2 $\mu$m), indicative of the presence of low-Ca pyroxene, would suggest that these objects formed from melting of ordinary chondrite material. This would make pallasites their 
most likely meteorite analogue. Our results suggest that R chondrites are better meteorite analogues for these asteroids. However, further analysis of meteorite samples will be needed to support or reject this idea.

\section{Dynamical connection with asteroid families}

\begin{figure*}[!ht]
\begin{center}
\hspace*{-1.5cm}
\includegraphics[height=9cm]{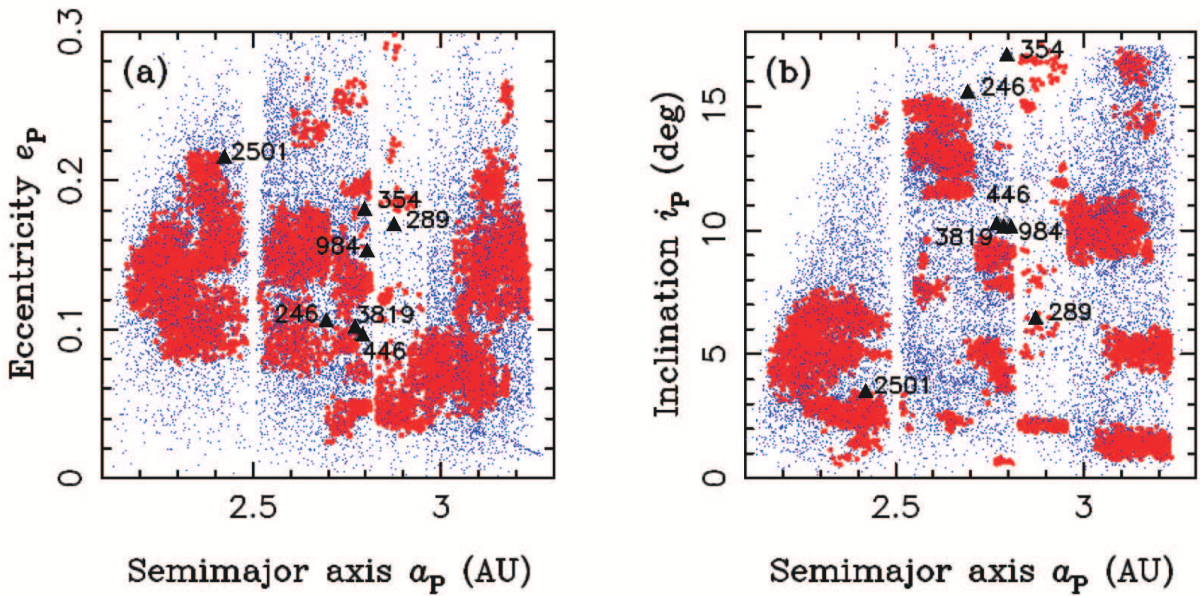}
\caption{\label{f:Dynamic} {\small Proper eccentricity ($e_{P}$) and proper inclination ($i_{P}$) vs. proper semimajor axis ($a_{P}$) for the asteroid families identified by \citet{2012PDSS..189.....N} (red), 
background objects (blue), and the S(I)-type asteroids: 246, 289, 354, 446, 984, 2501, and 3819 (filled triangles). Due to its high inclination, asteroid (863) Benkoela is not included in this figure.}}
\end{center}
\hspace*{-1.5cm}
\end{figure*}

In addition to the mineralogical analysis, an important part in understanding the formation of S(I)-type asteroids is to try to establish dynamical links between these objects and other asteroids or families. Some asteroid 
families show compositional diversity among their members. This diversity could be explained by the presence of interlopers in the family, but could also reflect the compositional gradient of what once was a differentiated 
asteroid later fragmented. Although conclusive evidence for the latter scenario still remains elusive, this hypothesis could not be ruled out in recent studies of asteroid families \citep[e.g.,][]{2008Icar..195..277M}. Thus, 
finding dynamical links between S(I)-type asteroids and metallic and/or basaltic asteroids (representing the core and crust of a differentiated body) could support the scenario in which these olivine-dominated asteroids 
originated from a differentiated object. Table \ref{t:Table5} includes the orbital elements and absolute magnitudes of the S(I)-type asteroids studied. After doing a careful inspection of the proper orbital elements 
($e_{P}$, $i_{P}$, $a_{P}$) of asteroids 246, 289, 354, 446, 863, 984, 2501, and 3819 we determined that none of these bodies are related to known families. In particular we compared the orbital elements of these 
asteroids with those from \citet{2012PDSS..189.....N}. The proper eccentricity ($e_{P}$) and proper inclination ($i_{P}$) versus proper semimajor axis ($a_{P}$) of these objects, along with the asteroid families identified 
by \citet{2012PDSS..189.....N}, and background objects are shown in Fig. \ref{f:Dynamic} as filled triangles, red, and blue points, respectively. This result is in agreement with the work of \citet{2013ApJ...770....7M}, in which 
they linked $\sim$ 38300 asteroids into 76 distinct families using the Hierarchical Clustering Method (HCM) and WISE/NEOWISE data. None of the S(I)-type asteroids listed above were linked to the dynamical families of 
\citet{2013ApJ...770....7M}.

\begin{table}[!ht]
\caption{\label{t:Table5} {\small Osculating elements and absolute magnitudes of the S(I)-type asteroids studied.}}
\begin{center}\small
\begin{tabular}{|c|c|c|c|c|}

\hline

Object&Semimajor axis (AU)&Eccentricity&Inclination ($^{o}$)&Absolute magnitude \\  \hline

246 Asporina&2.6975&0.1073&15.6265&8.62 \\
289 Nenetta&2.8719&0.2058&6.6972&9.51 \\
354 Eleonora&2.7985&0.1147&18.3925&6.44 \\
446 Aeternitas&2.7901&0.1261&10.6294&8.90 \\
863 Benkoela&3.1986&0.0347&25.3963&9.02 \\
984 Gretia&2.8032&0.1964&9.0860&9.03 \\
1951 Lick&1.3905&0.0616&39.0897&14.7 \\
2501 Lohja&2.4244&0.1945&3.3129& 12.08 \\
3819 Robinson&2.7734&0.1363&11.0977&11.8 \\
4125 Lew Allen&1.9214&0.1178&20.4399&13.5 \\
4490 Bambery&1.9311&0.0921&26.1157&12.7 \\
5261 Eureka&1.5236&0.0648&20.2810&16.1 \\

 \hline

\end{tabular}
\end{center}

\end{table}


\begin{figure*}[!ht]
\begin{center}
\includegraphics[height=10cm]{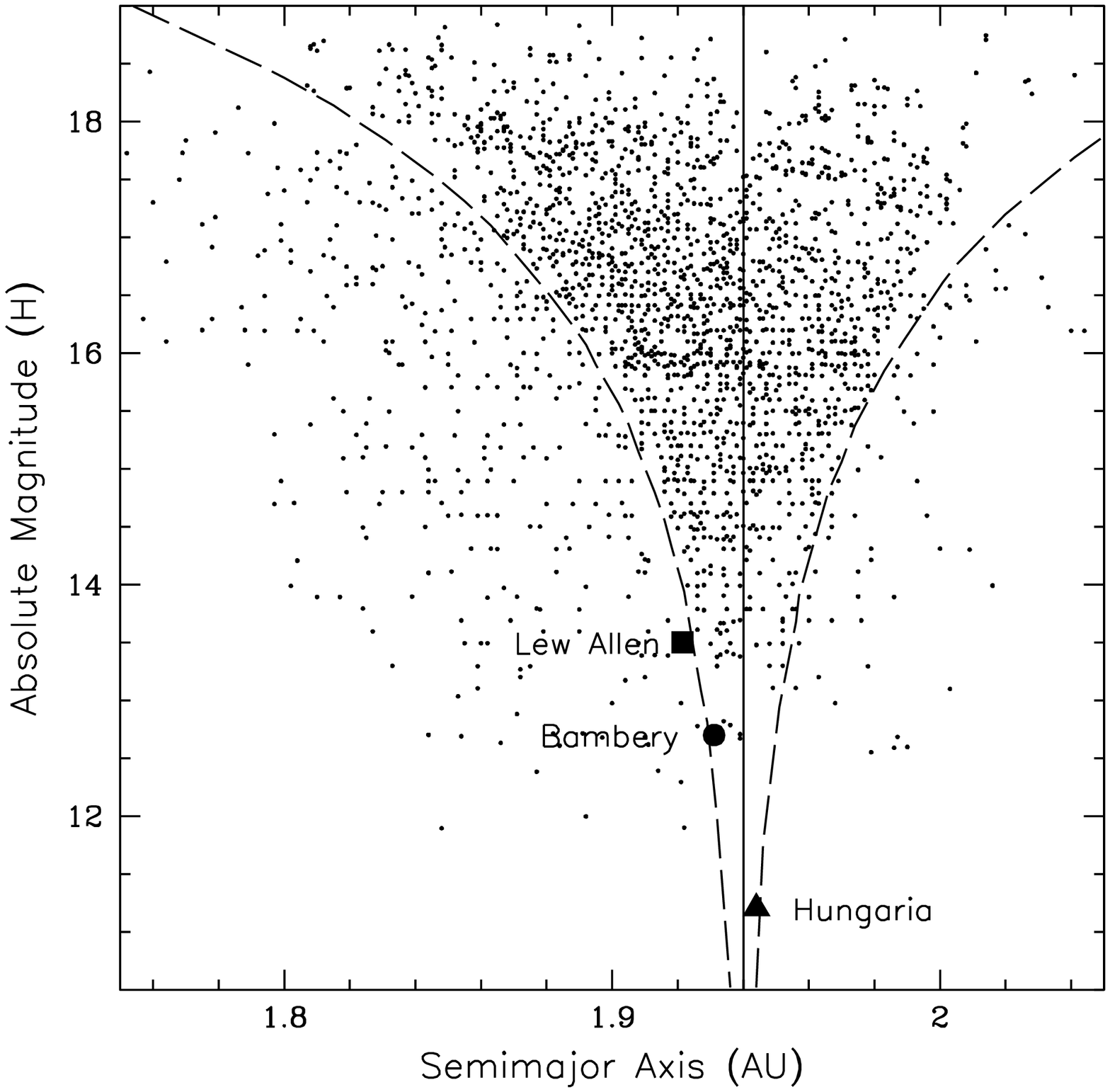}
\caption{\label{f:Ha} {\small Hungaria-population asteroids in a diagram of absolute magnitude (H) versus proper semimajor axis ($a_{P}$) from \citet{2009Icar..204..172W}. The dashed curved lines show the 
Hungaria-family region where the highest concentration of the asteroids is located. The vertical line represents the center of the collisionally-derived family \citep{2009Icar..204..172W}. The location of (434) Hungaria, 
(4125) Lew Allen, and (4490) Bambery are depicted as a filled triangle, filled square and filled circle, respectively.}}
\end{center}
\end{figure*}

Two of the studied asteroids, (4125) Lew Allen and (4490) Bambery, have orbital elements which place them in the same region of the Hungaria asteroid family, defined by a semimajor axis of 1.78 $<$ $a$ $<$ 2.0 
AU, an eccentricity $e < 0.18$, and inclination $16^{o}<i<34^{o}$ \citep{2009Icar..204..172W}. Figure \ref{f:Ha} shows the Hungaria family in a plot of absolute magnitude (H) versus proper semimajor axis ($a_{P}$)  from 
\citet{2009Icar..204..172W}. The location of (434) Hungaria, (4125) Lew Allen, and (4490) Bambery are depicted as a filled triangle, filled square and filled circle, respectively. Asteroid (434) Hungaria has been classified 
as E/Xe-type \citep{1984PhDT.........3T, 2009Icar..202..160D} and it has a high geometric albedo of $p_{v}=0.428$. The VIS-NIR spectrum of this asteroid exhibits a few weak absorption features in the 
visible region ($\sim$ 0.45-0.9 $\mu$m), and it is relatively smooth and featureless in the wavelength range of 0.8-2.5 $\mu$m \citep{2002M&PS...37.1815K}. A detailed analysis of (434) Hungaria 
carried out by \citet{2002M&PS...37.1815K} showed that the surface of this asteroid is mainly composed of iron-free pyroxene, enstatite, making enstatite achondrites (aubrites) the best meteorite analogs for this 
object. Similar results were obtained by \citet{2004JGRE..109.2001C}. The only way to produce an object of the size ($\sim$ 11 km) and composition of Hungaria is through the melting and differentiation of a parent body 
with an enstatite chondrite composition \citep{1989Metic..24..195K, 1999M&PS...34..735M, 2002M&PS...37.1815K}. However, this process would lead to the formation of an enstatite mantle instead of an olivine mantle. 
Therefore, if (434) Hungaria is a fragment of the enstatite mantle of a disrupted body (as suggested by the evidence), a genetic connection between this asteroid and (4125) Lew Allen and (4490) Bambery could be ruled 
out. Implying that these asteroids are probably interlopers in the Hungaria family.  


Asteroid (5261) Eureka is one of the eight Mars Trojans known to date (de la Fuente Marcos and de la Fuente Marcos, 2013). Seven of these objects (including Eureka) are located in the L$_{5}$ point and one of them in 
the L$_{4}$. Possible formation scenarios for Eureka and the Mars Trojans include: collisional fragmentation, rotational fission of the parent body, and capture events in the Trojan region 
\citep{2013Icar..224..144C, 2013MNRAS.432L..31D}. Only for three of these asteroids, 5261 Eureka (L$_{5}$ point), 101429 (1998 VF31) (L$_{5}$ point), and 121514 (1999 UJ7) (L$_{4}$ point), spectroscopic 
observations are available. VIS-NIR spectra ($\sim$0.4-2.5 $\mu$m) of asteroid 101429 (1998 VF31) obtained by \citet{2007Icar..192..434R} showed that this object belongs to the S(VII) subclass defined by 
\citet{1993Icar..106..573G}. Objects in this class are considered to be analogous to the mesosiderite meteorites (i.e., metal-basaltic assemblages). From their analysis, 
\citet{2007Icar..192..434R} determined that the spectrum of 101429 (1998 VF31) is consistent with a mixture of primitive achondrites and iron, and in general, their mixture models produced good fits with and without the 
contribution of mesosiderites. Based on these results, \citet{2007Icar..192..434R} concluded that there is no genetic connection between Eureka and 101429 (1998 VF31), and suggested that at least one of these objects 
was probably captured into its current orbit during the early stages of the solar system. These results are confirmed by recent numerical models carried out by \citet{2013MNRAS.432L..31D}. VIS spectra (0.5-1.0 $\mu$m) 
of asteroid 121514 (1999 UJ7) acquired by \citet{2003Icar..165..349R} allowed them to classify this object as an X-class \citep{1999PhDT........50B}. Spectra of this class are characterized by being featureless with 
moderately reddish slopes, and have been associated with iron meteorites and enstatite achondrites \citep{2002Icar..158..146B}. Unfortunately, the lack of NIR spectra for 121514 (1999 UJ7) prevents a more detailed 
mineralogical analysis. Based on the current data it is not possible to establish whether (5261) Eureka is indeed a fragment of the mantle of a primordial body, presumably formed in the terrestrial planets 
region, or whether it reached its current location during a capture event. Additional spectroscopic observation of the rest of the Mars Trojans could help to constrain the formation scenario for this object.  

Asteroid (1951) Lick has a perihelion distance of $q=1.305$ AU, which places it in the borderline between NEAs ($q\leq 1.3$ AU) and Mars-crossing asteroids ($1.3<q<1.66$ AU). Dynamical modeling performed 
by \citet{2013Icar..222..273D} showed that the source regions for olivine-dominated NEAs include: the intermediate source Mars-crossing region, the $\nu_{6}$ secular resonance, and the 3:1 mean-motion resonance 
with Jupiter. Apart from (1951) Lick, the only other olivine-dominated NEA with known composition is (136617) 1994CC \citep{2011P&SS...59..772R}. The analysis of this asteroid carried out 
by \citep{2011P&SS...59..772R} revealed a Mg-rich (Fo$_{90}$) olivine composition, similar to Mg-rich pallasites with a low metal component. This difference in olivine chemistry between (136617) 1994CC and 
(1951) Lick (Fo$_{70\pm5}$), along with the significant difference in their orbital elements suggest that there is no genetic connection between these two objects.


\section{Conclusions}

The surfaces of S(I)-type asteroids are dominated by olivine and in some cases pyroxene as a minor phase. Using VIS-NIR spectroscopy, mineralogy, and mineral chemistry of monomineralic-olivine asteroids can be 
constrained using olivine spectral calibration developed by \citet{1987JGR....9211457K} and improved by \citet{2011P&SS...59..772R}. We have found that the olivine composition for the monomineralic-olivine asteroids 
ranges from $\sim$ Fo$_{49}$ to Fo$_{70}$. Based on their olivine chemistry, albedo, and the absence of a significant 2 $\mu$m feature we determined that the most plausible meteorite analogues for (354) Eleonora, 
(1951) Lick and (4125) Lew Allen are brachinites. In the case of (246) Asporina, (289) Nenetta, (4490) Bambery and (5261) Eureka, brachinites are also the most likely meteorite analogues, although R chondrites can not 
be completely ruled out. 

For the olivine-rich asteroids we found similarities between their spectral band parameters and those measured for R chondrites. Therefore, we have developed a new set of spectral calibrations from the analysis of 
VIS-NIR spectra of R chondrites, whose modal abundances are known. We have established a relationship between the ol/(ol+px) ratio and BAR, and the LCP/(LCP+HCP) ratio and the Band II centers. Using these 
equations we have constrained the olivine and low-Ca pyroxene abundance of these objects. In particular, we found that the olivine abundance for the olivine-rich asteroids varies from 0.68 (863 Benkoela) to 0.93 (2501 
Lohja). The highest fraction of low-Ca pyroxene (0.9) was found for (446) Aeternitas, while the lowest (0.6) corresponds to (984) Gretia. Although this is a preliminary estimation, this is the first time that the mineral 
abundance of these asteroids has been determined. More laboratory measurements including both, spectra and modal abundances will be needed to improve the spectral calibrations. These additional data could also 
help to develop new equations to estimate the olivine and pyroxene chemistry of this type of asteroid. This would be particularly useful considering that the current calibrations 
\citep{1990JGR....95.6955S, 1987JGR....9211457K} are affected by the presence of pyroxene, and therefore can not be used when this mineral is present as a secondary phase. 

Based on the spectral and mineralogical characteristics of the studied asteroids, the two most plausible meteorite analogues are brachinites and R chondrites, which span the primitive-differentiated range. This suggests 
that olivine could be produced via nebular processes or in a thermally evolved environment. However, it is also possible that samples of S(I)-type asteroids have not yet been found in terrestrial meteorite collections.
 
A possible explanation for the paucity of olivine-dominated asteroids in the main belt is that all differentiated asteroids (with the exception of Vesta) were either disrupted or fragmented into pieces 
during the early stages of the solar system. These fragments, some from the crust and some from the mantle, were subsequently eroded by continuous collisions, being reduced to pieces that 
are below our current detection limits \citep[see][and references therein]{1996M&PS...31..607B}. Our failure in finding dynamical links between the S(I)-type asteroids and basaltic objects could support this hypothesis. On 
the other hand, depending on the starting material, the melting and differentiation of the parent body could form a mantle with a different composition. As discussed in the previous section, there is evidence that shows that 
differentiation of an enstatite chondrite-like body will produce an enstatite mantle instead of an olivine mantle.
 
Another possible scenario is that the largest S(I)-type asteroids were originally part of differentiated objects that started their lives outside the main belt.  In this circumstance, their parent body or bodies were disrupted by 
early solar system collisions, with the fragments scattered across the inner solar system. We speculate that this led a few S(I)-types to be captured within the primordial main belt by early dynamical processes.

In the existing literature, there are several possible ways S(I)-type asteroids (and other objects) may be captured within the primordial main belt.  First, it is possible these bodies reached their current orbits by being 
scattering into the primordial main belt via interacting planetary embryos \citep{2006Natur.439..821B}.  Second, the S(I)-types may have been injected into the main belt by the dynamical conditions surrounding the 
so-called "Grand Tack" \citep{2011Natur.475..206W}.  In this scenario, Jupiter migrates both inward and outward across the primordial main belt via interactions with the solar nebula. This may have allowed objects in the 
inner solar system to become deeply embedded within the primordial main belt. A third possibility is that the S(I)-types were captured within the fossil resonances of Jupiter before the so-called Nice model events took 
place \citep{2013LPICo1719.1672B}. Fourth, it is plausible they were somehow captured during the events surrounding the Nice model itself \citep{2005Natur.435..459T}. This last mechanism seems less likely to us as of 
this writing, however, because the resonances that sweep across the main belt during the Nice model seem more likely to deliver comet-like bodies to the outer main belt than fragments from differentiated bodies 
\citep{2009Natur.460..364L}.

Regardless of the exact mechanism, if one or more of these capture scenarios are true in the broadest sense, it seems likely that the S(I)-types were once part of differentiated planetesimals that started their lives in the 
terrestrial planet region \citep{2006Natur.439..821B}.  This makes their origin and evolution history extraordinarily interesting from the perspective of understanding terrestrial planet formation.



\

{\bf{Acknowledgements}}

\
  

This paper is based on data obtained with the Infrared Telescope Facility on Mauna Kea, Hawai'i. Some of the data used in this work were obtained from the SMASS II and S$^{3}$OS$^{2}$. In addition, part of the data 
utilized in this publication were obtained and made available by the MIT-UH-IRTF Joint Campaign for NEO Reconnaissance. The IRTF is operated by the University of Hawaii under Cooperative Agreement no. NCC 
5-538 with the National Aeronautics and Space Administration, Office of Space Science, Planetary Astronomy Program. The MIT component of this work is supported by NASA grant 09-NEOO009-0001, and by the National 
Science Foundation under Grants Nos. 0506716 and 0907766. Any opinions, findings, and conclusions or recommendations expressed in this material are those of the author(s) and do not necessarily reflect the views of 
NASA or the National Science Foundation. This publication makes use of data products from the Wide-field Infrared Survey Explorer, which is a joint project of the University of California, Los Angeles, and the Jet 
Propulsion Laboratory/California Institute of Technology, funded by the National Aeronautics and Space Administration. This publication also makes use of data products from NEOWISE, which is a project of the Jet 
Propulsion Laboratory/California Institute of Technology, funded by the Planetary Science Division of the National Aeronautics and Space Administration. The authors thank, John Hinrichs and Paul Lucey for providing us 
with data for this research. VR and MJG research was supported by NASA 182 NEOO Program Grant NNX12AG12G, and NASA Planetary Geology and Geophysics Grant NNX11AN84G. We thank the IRTF TAC for 
awarding time to this project, and to the IRTF TOs and MKSS staff for their support. We also thank the anonymous reviewers for their useful comments, which helped to improve the manuscript.

\bibliographystyle{model2-names}
\bibliography{references}







\end{document}